\documentclass{article}

\usepackage{graphicx}
\usepackage{epstopdf,epsfig}
\usepackage{natbib}
\usepackage{hyperref}
\hypersetup{
    colorlinks = true,
    urlcolor   = blue,
    citecolor  = black,
}

\newcommand{\RomanNumeralCaps}[1]
\linenumbers

\usepackage[percent]{overpic}
\usepackage{mathtools}
\usepackage{stackengine}
\def \BEA {\begin{eqnarray}} 
\def \EEA {\end{eqnarray}}
\def \CR {\nonumber \\}

\def \p {{\textbf p}}
\def \r {{\textbf r}}

\def \P {\cal P}
\def \q {{\textbf q}}

\def \K {{\textbf k}}

\def \x {{\textbf x}}

\def \OMEGAP {\sigma_\p}

\def \H {{\cal H}}

\newcommand{\sizeA}{\rule{0mm}{6mm}}

\newcommand{\LSBA}{\mbox{$\left[\sizeA\right.$}}  

\newcommand{\RSBA}{\mbox{$\left.\sizeA\right]$}}  

\begin{document}
\title{Generalized Transport Characterizations for Short Oceanic Internal Waves in a Sea of Long Waves}

\author{Yuri V. Lvov$^1$ and   Kurt L. Polzin$^2$\\
$^1$
Department of Mathematical Sciences, Rensselaer Polytechnic Institute, Troy, NY\\
$^2$
Woods Hole Oceanographic Institution, MS\#21, Woods Hole, MA 02543}

\maketitle

\begin{abstract}
  
  Internal waves in the ocean interact in triads. Early work
  emphasized the importance of extreme-scale separated interactions in
  which two large wavenumber waves interact with one small wavenumber
  wave. More recent efforts have called this early paradigm into
  question.  We use wave turbulence kinetic equation and the
  ray-tracing WKB technique to derive two versions of the
  corresponding Fokker-Planck (generalized diffusion) equation. We
  then use these Fokker-Planck equations to estimate the spectral
  energy flux towards dissipation (high wave numbers) we obtain
  different results: spectral transport of the kinetic equation
  Fokker-Planck equation is an order of magnitude larger than either
  observations or reported ray tracing estimates.  This apparent
  contradiction stems from the difference between Eulerian and
  ray-path descriptions of these scale-separated interactions.

\end{abstract}

\section{Introduction}
Internal waves are a fascinating phenomenon, ubiquitous in the ocean, and characterized by the oscillation of the invisible surfaces of constant densities of a stratified water column.  Internal waves carry a significant fraction of ocean kinetic energy and are an important intermediary in transferring energy and momentum to smaller scales where they are dissipated.  There is renewed theoretical interest for investigating the internal waves in the ocean due to recent developments in our ability to perform high-resolution numerical modeling, with internal wave permitting Global Ocean Simulations \citep{arbic2018primer} and regional numerical models \citep{pan2020numerical}.

The wave turbulence kinetic equation has been used extensively to describe processes of spectral energy transfers between internal waves, see  \cite{M86,Regional} for reviews.  Early on, special types of resonant three-wave triads characterized by extreme scale separations were identified to play an important role in these spectral energy transfers \citep{MB77,MMa}, yet the details and delicate interbalance of the nonlinear transfers remain an enigma. An important feature of the internal wave kinetic equation is that it diverges for almost all spectral power law indices in the internal wave spectrum \citep{theory}. This is a mathematical manifestation of the lack of locality in internal wave interactions: nonlinear transfers have significant contributions under extreme scale-separated conditions.  The weighting of these extreme scale-separated interactions with spectral power law assumptions results in divergent integrals.  Our goal is to understand and to characterize this nonlinear transfer process in this extreme scale-separated limit.

An alternative approach to the wave turbulence kinetic equation is proposed in \cite{HWF}, where ray-tracing (or eikonal) techniques are used to describe spectral energy transfers.  In this paradigm, the energy cascade is assessed as a net drift $\langle \dot {\bf p} \rangle$ of wave packets toward high wavenumber, where ${\bf p}$ is the momentum of a wave and $\langle\dots\rangle$ represents an ensemble average. Such studies (\cite{HWF,SK99b,ijichi2017eikonal}) provide metrics of the net drift rate at a high wavenumber gate, beyond which waves are considered to 'break'.  These numerical simulations are conducted in a 'kitchen sink' manner in which scale separations in vertical and horizontal wavenumber are viewed as tunable parameters to arrive at downscale transport estimates that align with observations.  The alignment requires that the background have similar scales as the wave packet and creates a thematic issue for an asymptotic theory such as ray tracing.  A further issue is that a rigorous description of this ensemble average transport $\langle \dot{\p} \rangle$ is an open question that we address here.  
    
Motivation for our efforts comes from a comparison with the empirical metrics of ocean mixing referred to as 'finescale parameterizations' \cite{G89,P95}, see \cite{Finescale} for a review and \cite{M77,MMb,Ultraviolet,dematteis2022origins} for descriptions of the pivotal role that extreme scale separated interactions play in interpreting the oceanic internal wave spectrum.  The most glaring incompatibility of wave turbulence and ray tracing is presented in Section \ref{sec:EnergyTransport}:  If the mean drift rate in vertical wavenumber is identified as the corresponding gradient of diffusivity as derived from the kinetic equation, then the predicted downscale energy transport is an order of magnitude larger than that supported by the observations.  This method parallels assessments for downscale energy transport in ray-tracing numerics \citep{HWF,SK99b,ijichi2017eikonal}, but is similarly ten times larger than those numerical results.  This disparity has lead us to a systematic examination and physical interpretation of the assumptions within both kinetic equation and ray-path approaches to pinpoint the multiple junctures which might underpin a systematic difference between observation and theory concerning extreme scale separated interactions.  

Despite claims by \cite{MB77} and \cite{SnazarNonlocal1} that ray tracing should reduce to the resonant manifold, our understanding is that the kinetic equation and ray tracing differ on fundamental levels.  The wave kinetic equation represents the internal wavefield as a system of amplitude modulated waves having constant wavenumber and frequency linked through a dispersion relation.  Ray tracing represents a wave-packet as a frequency-modulated system with variations in wavenumber linked to the conservation of an Eulerian phase function.  The Fokker-Plank derived in the ray tracing approach additionally represents the average drift of wave packets towards high wave numbers.  The role of resonances and off-resonant interactions in the mean drift and dispersion about that mean drift are also different \citep{polzinCompanion}, as are the concepts of resonance broadening \citep{Ultraviolet} and bandwidth \citep{CohenLee} that are important metrics of finite amplitude effects in weakly nonlinear systems.

Our efforts have direct parallels with Kraichnan's 1959 and 1965 studies \citep{kraichnan1959structure,kraichnan1965lagrangian} of isotropic homogeneous turbulence using field theoretic techniques.  Kraichnan's 1959 study was an Eulerian based approach which yielded a $k^{-3/2}$ spectrum at high Reynolds number, distinct from Kolmogorov's $k^{-5/3}$ inertial subrange based upon dimensional analysis.  This Eulerian description was labeled the 'Direct Interaction Approximation' (DIA) and extant data were not sufficient to assess the theoretical prediction.  \cite{kraichnan1965lagrangian} subsequently understood that the quasi-uniform translation associated with coherent advection at the largest scales ({\it aka} sweeping) was creating an artifact wherein the correlation time scale was proportional to the root-mean-square Doppler shift rather than a more intuitive notion that energy transfers between scales depended upon the rate of strain.  In 1965 Kraichnan subsequently presented a Lagrangian description (the Abridged Lagrangian History DIA) that isolated the pressure and viscous terms responsible for fluid parcel deformation.  The Lagrangian picture resulted in a -5/3 power law and Kolmogorov constant (1.77) quite close to that provided by a summary of atmospheric field data (1.56) (\cite{hogstrom1996review}).  The analogy to Kraichnan is that the plane wave formulation corresponds to an Eulerian coordinate system and the Lagrangian coordinate system corresponds to a wave packet formulation in which statistics are accumulated along {\it ray paths}.  

Similar issues about Doppler shifting arise for internal waves \citep{Holloway1,Holloway2}, Rossby waves \citep{holloway1977stochastic, nazarenko2011wave} and in Magneto-HydroDynamics \citep{SnazarNonlocal1}.  Wave problems are potentially more complicated, in part because the Doppler shifting can be intrinsically related to extreme scale separated interactions, and due to a multiplicity of time scales introduced through resonant interactions absent in 3-D turbulence.  In wave turbulence one assumes an expansion in terms of small nonlinearity and an assumption about multiple time scales to assess the evolution of amplitude modulated plane waves.  Implicit is a long interaction time scale and a short time scale with regards to the higher orders \citep{newell1968closure}.  Reduction of the DIA to the resonant manifold happens as the correlation time scale is small relative to an interaction time scale, and triple correlations associated with nonlinear coupling can be related to the product of two double correlations.  Extreme scale separated wave problems can also be treated with ray tracing methods, in which statistics of frequency modulated wave packets are accumulated along ray characteristics rather than Lagrangian trajectories.  Ray tracing is an extremely attractive route to deal with sweeping as the dynamics of ray tracing are grounded in the explicit representation of variations in Doppler shifting.  It is understood that there are ray method parallels to the interaction and correlation time scales of wave turbulence and the DIA, \citep{MB77,SnazarNonlocal1}.  However, the time scale definitions for ray methods have not been sufficiently developed for a detailed comparison of the two strategies for assessing the effects of Doppler shifting.  In particular, what has been missing is the identification of the interaction time scale.  Here we provide a derivation of a generalized transport equation for the evolution of an ensemble of wave packets.  This generalized transport equation contains a term representing the ensemble mean drift of wave packets in the spectral domain.  This mean drift  relates to the interaction time scale and can be directly compared to a correlation time scale relating to dispersion about that mean drift.  Having accomplished this, we arrive at the understanding that the resonant bandwidths of weakly nonlinear interactions in the two systems, the DIA kinetic equation and from ray methods, are different; that resonant and non-resonant interactions express themselves differently in the correlation time scale than previously understood; and that spectral transports can be significantly altered by the mean-drift term.

We demonstrate here that it is this simple difference in coordinate systems that leads to the celebrated Garrett and Munk spectrum \footnote{We utilize what is referred to as GM76, the 1976 version of their model.  We refer the reader to \cite{GM72,GM79} for historical perspectives and to \cite{M86,Regional} for reviews.} of the oceanic internal wavefield supporting a net downscale transport.  The 3-d action spectrum for the Garrett and Munk spectrum is independent of vertical wavenumber, so that in an Eulerian description there is no vertical wavenumber action gradient to support the diffusion of action regardless of how the vertical component of the diffusivity tensor is defined.  In a ray description, the mean drift of wave packets to a high wavenumber can be explicitly represented in an enesmble transport equation and an estimate of the action (energy) available for mixing can be obtained by the counting of wave packets past a sufficiently high wavenumber gate \citep[e.g][]{HWF}.

This paper is organized as follows. Hamiltonian structures and the 
derivation of transport equations from them are the focal points of 
Sections \ref{Background} and \ref{ScaleSeparations}.  We review the 
Hamiltonian structure in section \ref{HamiltonianSection}.  
In Section \ref{WTSection} we 
present a derivation for internal waves that leads to a Fokker-Planck equation.  In Section \ref{ScaleSeparations} we refine the Hamiltonian structure; extracting only those extreme scale separated interactions in order to derive the Liouville equation (Section \ref{S1}) and its subsequent Fokker-Planck (Section \ref{WKBDiffusion}).  Subsequent to these theoretical developments we present estimates of energy transport to mixing scales and demonstrate the mismatch between theory and observations in Section \ref{sec:EnergyTransport}.  We end in Section \ref{sec:Conclusions} by discussing this contradiction in light of our derivations.  The reader who is primarily interested in the disparity between observations and theory is advised to read section \ref{sec:EnergyTransport} and use the equation references to navigate Sections \ref{Background} and \ref{ScaleSeparations}.  

\section{Background}\label{Background}
\subsection{Hamiltonian Structure and Field Variables}
\label{HamiltonianSection}

The equations of motion satisfied by an incompressible stratified rotating flow in hydrostatic balance are

\begin{eqnarray}
\frac{\partial}{\partial t}\frac{\partial z}{\partial \rho} + \nabla \cdot \left(\frac{\partial z}{\partial \rho} \bf{u} \right) &=& 0 , \nonumber \\
\frac{\partial \bf{u}}{\partial t} +f \bf{u}^\perp+ \bf{u} \cdot
\nabla \bf{u} + \frac{\nabla M}{\rho} &=& 0 ,
\nonumber
\\
\frac{\partial M}{\partial \rho} - g z &=& 0 .
\label{PrimitiveEquations}
\end{eqnarray}

These equations result from mass conservation, horizontal momentum
conservation and hydrostatic balance.  The equations are written in
isopycnal coordinates with the density $\rho$ replacing the height $z$
in its role as an independent vertical variable.  Here ${\bf u} = (u,v)$
is the horizontal component of the velocity field, ${\bf u}^{\perp} =
(-v, u)$, $\nabla = (\partial/\partial x, \partial/\partial y)$ is the
gradient operator along isopycnals, $M$ is the Montgomery potential
$$M=P+g\,\rho\,z \, , $$
with pressure $P$, gravity $g$ and Coriolis parameter $f$.  

Here we follow \cite{LT2} and take equations
(\ref{PrimitiveEquations}) and decompose the flow into a potential and
a divergence-free part:
\begin{equation}
  \bf{u} = \nabla \phi + \nabla^{\perp} \psi \, ,
\label{decomp}
\end{equation}
where 
\begin{equation}
  \nabla^{\perp} =\left( \begin{array}{r}
        -\partial_y \\ 
         \partial_x 
         \end{array} \right) \, .
\end{equation} 

The expression for potential vorticity in these coordinates is \cite{haynes1987evolution}
\begin{equation}
{\cal Q} = \frac{f+\partial v/\partial x - \partial u/\partial y}{\Pi} ,
\label{PotentialVorticity}
\end{equation}
where
$\Pi = \frac{\rho}{g} \partial^2 M/\partial \rho^2 = \rho \partial z/\partial \rho $
is a normalized differential layer thickness. Since potential vorticity is conserved along particle trajectories, 
\begin{equation}
 \frac{D {\cal Q}}{D t} = 0 . \label{Vorticity}
\end{equation}
The advection of potential vorticity in (\ref{Vorticity})
takes place exclusively along isopycnal surfaces. Therefore, an
initial distribution of potential vorticity which is constant on
isopycnals, though varying across them, will remain constant.  Hence we
shall utilize 
\begin{equation}
  {\cal Q}(x, y,\rho,t)  = f/\Pi_{0},
 \label{PV}
\end{equation}
where we redefined $\Pi \to \Pi_{0}+\Pi $ to
split the potential $\Pi$ into its equilibrium value
$\Pi_{0} \equiv - g / N^2$ and deviation from it.  Stratification $N^2$ is permitted to vary with density $\rho$, but is constant along isopycnals.  This effectively decouples the internal wavefield from lower frequency flows such as fronts and mesoscale eddies, which are the subject of their own wave turbulence literature \citep[e.g.][]{M76,kafiabad2019diffusion}.  Such internal wave - mean flow interactions can be a significant regional source of internal wave energy \citep{polzin2010mesoscale} and may be a key issue in determining the regional character of the internal wavefield \citep{Regional}.  

The primitive equations of motion
(\ref{PrimitiveEquations}) under the assumption ({\ref{PV}) can be
  written as a pair of canonical Hamiltonian equations,
\begin{equation}
 \frac{\partial \Pi}{\partial t}  =  \frac{\delta {\cal H}}{\delta \phi} \, ,
  \qquad
\frac{\partial \phi}{\partial t}  = -\frac{\delta {\cal H}}{\delta \Pi} \, ,
  \label{Canonical}
\end{equation}
where $\phi$ is the isopycnal velocity potential, and the Hamiltonian is the sum of kinetic and potential energies,
%
{
\begin{eqnarray}
 {\cal H} = \!\! \frac{1}{2}\int \!\! d \x d\rho
\left[
 \left(
\Pi_{0}(\rho)+\Pi({\bf{x}}, \rho) \right) \,
\left|\nabla \phi(\x,\rho) + \frac{f}{\Pi_{0}}  
\nabla^{{\perp}}\Delta^{-1} \Pi(\bf{x}, \rho)
  \right|^2
-g \left|\int^{\rho} d\hat{\rho} \frac{\Pi(\x,\hat{\rho})}{\hat{\rho}}  
      \right|^2
  \right]
\nonumber\\
  \label{HTL2}
\end{eqnarray}
%
with $\nabla^{\perp}=(-\partial / \partial y, \partial / \partial x)$, $\Delta^{-1}$ is the inverse Laplacian and $\hat{\rho}$
represents a variable of integration.

Our intent is to build a perturbation theory around analytical solutions to the linearized primitive equations as plane waves proportional to $e^{i [\r \cdot \p - \sigma t]}$. 
We therefore transition to the Fourier space:
\begin{eqnarray}
\Pi(x, y, \rho) = \frac{1}{(2 \pi)^{3/2}}\int \Pi_{\bf p} \; e^{ i  {\bf r} \cdot {\bf p}} d {\bf p}, & &
\phi(x, y, \rho) = \frac{1}{(2 \pi)^{3/2}}\int \phi_{\bf p} \; e^{ i {\bf r} \cdot {\bf p}} d{\bf p},  \nonumber  \\
{\bf p}=( {\bf{k}},m), & 
{\bf k}=( k, l), &
{\bf r}=(x, y, \rho) \; , 
\label{Fourier}
\end{eqnarray}
and introduce a complex field variable $a_{\bf{p}}$ through the canonical transformation
\begin{eqnarray}
 \phi_{\bf{p}} = \frac{i N \sqrt{\sigma_{\bf{p}}}}{\sqrt{2 g} |\bf{k}|} \left(a_{\bf{p}}-
a^{\ast}_{-{\bf{p}}}\right)\, , \ \ \ 
\Pi_{\bf{p}} = \frac{\sqrt{g} |\bf{k}|}{\sqrt{2\sigma_{\bf p}}N}\left(a_{\bf{p}}+a^{\ast}_{-{\bf{p}}}\right)\, .  
\label{TransformationToActionVariables}
\end{eqnarray}
Wave frequency $\sigma_{\bf{p}}$ is restricted to be positive.  We
ignore variations in density as they multiply horizontal momentum,
replacing $\rho$ by a reference density $\rho_0$ (the Boussinesq
approximation) and arrive at a linear dispersion frequency $\sigma$
given by 
\begin{eqnarray}
\sigma_{\bf{p}}=\sqrt{ f^2 + \frac{g^2}{\rho_0^2 N^2} \frac{|\bf{k}|^2 }{m^2}} \; .
\label{InternalWavesDispersion}
\end{eqnarray}
The equations of motion (\ref{PrimitiveEquations}) adopt the canonical form
\begin{equation}
i\frac{\partial}{\partial t} a_{\bf{p}} = \frac{\delta {\cal H}}{\delta
a_{\bf{p}}^{\ast}} \, ,\label{fieldequation}
\end{equation}
with Hamiltonian:
\begin{eqnarray}
&& {\cal H} = \int d\bf{p} \, \sigma_{\bf{p}} |a_{\bf{p}}|^2
\nonumber \\
 && 
\quad
+ \int d{\bf{p}} d {\bf p_1} d {\bf p_2} 
\left(
 \delta_{\bf{p}+\bf{p}_1+\bf{p}_2} (U_{\bf{p},\bf{p}_1,\bf{p}_2} a_{\bf{p}}^{\ast} a_{\bf{p}_1}^{\ast} a_{\bf{p}_2}^{\ast} + \mathrm{c.c.})
+ \delta_{-\bf{p}+\bf{p}_1+\bf{p}_2} (V_{\bf{p}_1,\bf{p}_2}^{\bf{p}} a_{\bf{p}}^{\ast} a_{\bf{p}_1} a_{\bf{p}_2} + \mathrm{c.c.})
\right) . \nonumber\\
\label{HAM}
\end{eqnarray}
Here $V_{\bf{p}_1,\bf{p}_2}$ and $U_{\bf{p}_1,\bf{p}_2}$ are the
interaction cross sections that define the strength of nonlinear
interactions between wave numbers ${\bf p}$, ${\bf p_1}$ and ${\bf
  p_2}$~\cite{LT}; c.c. denotes the complex conjugate.  Implicit in
the canonical transformation (\ref{TransformationToActionVariables}),
Hamilton's equation (\ref{fieldequation}) and Hamiltonian (\ref{HAM})
is a time dependence of $e^{-i \sigma t}$.  The $U$ elements have a time dependence of 
$e^{i(\sigma_{\p_1} + \sigma_{\p_2} + \sigma_{\p_3})t}$ with $\sigma_{\p} >
0$.  They describe the creation of three waves out of nothing and
therefore will not appear in the kinetic equation (\ref{BroadenedKineticEquation}).

This is the standard form of the Hamiltonian of a system dominated by
three-wave interactions~\cite{ZLF}.  Calculations of interaction
coefficients are tedious but straightforward task, completed in
\cite{LT2,theory}.
We stress that the field equation (\ref{fieldequation}) with the
three-wave Hamiltonian (\ref{InternalWavesDispersion}, \ref{HAM}) is
{\em equivalent\/} to the primitive equations of motion for internal
waves (\ref{PrimitiveEquations}) with the potential vorticity
constraint (\ref{PV}).

\subsection{Wave Turbulence theory}
\label{WTSection}
In wave turbulence theory, one proposes a perturbation expansion in
the amplitude of the nonlinearity, yielding linear waves at the
leading order. Wave amplitudes are modulated by the nonlinear
interactions, and the modulation is statistically described by a
kinetic equation~\citep{ZLF, nazarenko2011wave} for the wave action spectral density 
$n_{\bf{p}}$ defined by
\begin{eqnarray}
n_{\bf{p}} \delta(\bf{p} - \bf{p}^{\prime}) = \langle a_{\bf{p}}^{\ast} a_{\bf{p}^{\prime}}\rangle.
\label{WaveAction}
\end{eqnarray}
Here $ \langle \dots \rangle$ denotes an ensemble averaging,
i.e. averaging over many realizations of the random wave field.
Application to the internal wave problem is presented in Section 2b of
\cite{theory}.
\begin{figure}
%

\includegraphics[width=0.5\textwidth]{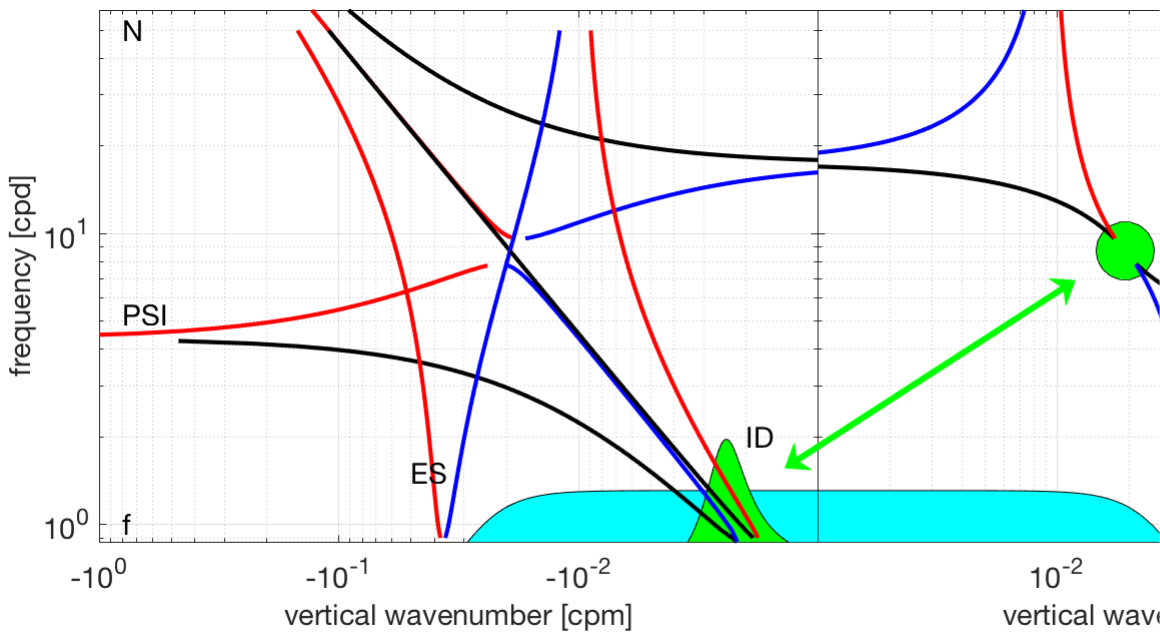}
\caption{The resonant manifold (\ref{ResonantManifold}) in the situation where the three horizontal wavevectors are either parallel or anti-parallel, plotted in a vertical wavenumber - frequency space, for a wave at the center of the green circle.  With rotation, extreme scale separations in horizontal wavenumber lead to the extreme scale separated triads mentioned in the introduction. These triads are  Bragg scattering (also called {\it elastic scattering, or} ES) and a phase velocity equals ground velocity resonance condition, called {\it Induced Diffusion, or} (ID) being located at the Coriolis frequency $f$.  This study focuses upon the latter class, with scale separation in both horizontal and vertical wavenumber.  Near-resonant ID conditions are depicted in green, bandwidth limited non-resonant ID forcing in cyan.  The third type of extreme scale separated triads, called {\it Parametric Subharmonic Instability} or PSI does not play a role in this manuscript.} \label{Triads}
\end{figure}

\subsubsection{Generalized (Broadened) Kinetic Equation: \label{DIA}}
In the limit of small nonlinearity, one develops a perturbation expansion in the nonlinearity strengh, which leads under certain assumptions to the wave turbulence kinetic equation. The derivation of the resonant kinetic equation is well understood and studied, see \cite{ZLF,nazarenko2011wave}.  Taking nonresonant interactions leads to a different version of the kinetic equation with the frequency delta functions being replaced by a Lorenzian, see \cite{LLNZ,element}.  This derivation also hinges on the assessment that fourth
order cumulants are a subleading term compared to the product of two double
correlators \citep{deng2021full}. 
For the three-wave Hamiltonian (\ref{HAM}),
the kinetic equation is Eq.~(\ref{BroadenedKineticEquation}), describing general internal
waves interacting in both rotating and non-rotating environments:
\begin{eqnarray}
\frac{\partial}{\partial t} n_{\p} =  \displaystyle \int
\int d\p_1 d\p_2 & \Big( \mid V_{\p_1,\p_2}^{\p} \mid^2
\delta(\p-\p_1-\p_2){\mathcal L}
(\Delta \sigma_{p12}, \Gamma_{p12})
      [n_{\p_1}n_{\p_2}-n_{\p}[n_{\p_1}+n_{\p_2}]]
      \nonumber
      \\ & \left.- \mid
      V_{\p_2,\p}^{\p_1} \mid^2 \delta(\p-\p_1+\p_2)
      {\mathcal
        L}(\Delta \sigma_{12p}, \Gamma_{p12})
      [n_{\p_2}n_{\p}-n_{\p_1}[n_{\p_2}+n_{\p}]]
      \right.
      \nonumber
      \\
            & - \mid
      V_{\p,\p_1}^{\p_2} \mid^2 \delta(\p+\p_1-\p_2){\mathcal
        L}(\Delta \sigma_{2p1}, \Gamma_{p12})
      [n_{\p}n_{\p_1}-n_{\p_2}[n_{\p}+n_{\p_1}]] \Big) .  \nonumber \\
\label{BroadenedKineticEquation}
\end{eqnarray}
where Laurencian $\mathcal{L}$ is given by
%
${\mathcal L} = \frac{\Gamma_{p12}}{(\Delta \sigma)^2 + \Gamma_{p12}^2},
$
and
$\Delta\sigma_{p12}=\sigma_p-\sigma_{p_1}-\sigma_{p_2}$
represents the distance from the resonant surface.  The resonant manifold is defined by
\begin{eqnarray}\label{ResonantManifold}
\sigma & = & \sigma_1 + \sigma_2 ; \;\;\;\;\; \;\;\;\;\;\; \bf{p} = \bf{p}_1 + \bf{p}_2  \nonumber \\
\sigma & = &  \sigma_1 - \sigma_2 ; \;\;\;\;\;\;\;\; \bf{p} =  \bf{p}_1-  \bf{p}_2   \nonumber \\
\sigma & = &  \sigma_2 - \sigma_1 ; \;\;\;\;\;\; \bf{p} = \bf{p}_2-  \bf{p}_1   \
\end{eqnarray}
and appears in figure \ref{Triads}.  In wave turbulence theory of
internal waves the importance of special extreme scale separated
triads was recognized in \cite{MB77}. These extreme-scale separated
limits are called Induced Diffusion (ID), Elastic Scattering (ES) and
Parametric Subharmonic Instability (PSI). For an explanation of these
triads see, as well, \cite{MMa}.

The total resonance width associated with a specific triad is given by
$
\Gamma_{p12} = \gamma_p + \gamma_1 +\gamma_2,
$
and the equation for the individual resonance widths is given by
\begin{eqnarray}
\gamma_p =  \displaystyle \int \int d\p_1 d\p_2
&\Big(  \mid V_{\p_1,\p_2}^{\p} \mid^2
\delta(\p-\p_{1}-\p_2) {\mathcal L} (\sigma-\sigma_1-\sigma_2) [n_{\p_1}+n_{\p_2}]
\nonumber \\
& \left.
+ \mid V_{\p_2,\p}^{\p_1} \mid^2 \delta(\p-\p_1+\p_2){\mathcal L}(\sigma-\sigma_1+\sigma_2)
 [n_{\p_2}-n_{\p_1}]\right. \nonumber \\ 
 &
+ \mid V_{\p,\p_1}^{\p_2} \mid^2 \delta(\p+\p_1-\p_2){\mathcal L}(\sigma+\sigma_1-\sigma_2)
 [n_{\p_1}-n_{\p_2}]
\Big).\nonumber\\
\label{ResonanceWidth}
\end{eqnarray}

Physically this $\gamma_p$ represents the fast time scale of decay of a narrow perturbation to the otherwise stationary spectrum \citep{LLNZ,Ultraviolet}.  It coincides with Langevin rates estimated by \cite{PMW80} and the decay rate of \cite{M77}'s spike experiments.  The replacement of the frequency conserving delta function by the Lorentzian takes into account not only resonant, but also near-resonant and nonresonant interactions. Nonresonant interactions appear as a result of the Lorentzian decaying slowly. The role of the nonresonant interactions have to be investigated separately for each particular problem. The detailed investigation that will be presented elsewhere,  show that for the case of internal waves and the Garrett and Munk spectrum, the nonresonant interactions leads to the Doppler defect, noticed in \cite{Ultraviolet}.

\subsubsection{Fokker-Planck Diffusion Limit}
Following \cite{MB77} for the resonant kinetic equation, we start from (\ref{BroadenedKineticEquation}) and pick off the
interactions having $\p$ nearly parallel to $\p_1$ with $\p_2$ small,
or nearly parallel to $\p_2$ with $\p_1$ small , which selects the ID
class triads.  For a sufficiently red spectrum, this permits
discarding the small $n_\p n_{\p1}$ ($n_\p n_{\p2}$, respectively)
terms.  We then rewrite (\ref{BroadenedKineticEquation}) as
\begin{equation}
\frac{\partial n_{\p}}{\partial t} = \int d \q \left( {\cal B}(\p)-{\cal
  B}(\p+\q)\right)\simeq -\int d
\q\ \ \left(\q\cdot\frac{\partial}{\partial \p}\right) {\cal B}(\p),
\label{Equation23}
\end{equation}
where we introduced
$$
{\cal B}(\p) =  8\pi \int
 |V_{\p_1,\q}^{\p}|^2 \, n_q(n_{\p1}-n_\p) \,
\delta_{{\p - \p_1-\q}} \, {\cal L}({\sigma_{\p}
-\sigma_{{\p_1}}-\sigma_{{\q}}})
d \p_{1}, 
$$ and expanded the difference $\left( {\cal B}(\p)-{\cal
  B}(\p+\q)\right)$ in a Taylor series using $\q$ to represent the
small difference in wavenumber between the two high frequency waves.
Expanding the difference $n_{\p_1}-n_\p$ for small $\q$ gives
\BEA\label{Equation24} {\cal B}(\p) \simeq -8\pi
\left(\q\cdot\frac{\partial n_\p}{\partial \p}\right) \, \int
|V_{\p_1,\q}^{\p}|^2 \, n_q \delta_{{\p - \p_1-\q}} \, {\cal
  L}({\sigma_{\p} -\sigma_{{\p_1}}-\sigma_{{\q}}}) d \p_{1}.\EEA
Combining (\ref{Equation23}) with (\ref{Equation24}) we obtain
\BEA
\frac{\partial  n_{\p}}{\partial t} & = & 
\frac{\partial}{\partial p_i} D_{ij}(\p,\q) \frac{\partial}{\partial p_j} n_{\p},
\nonumber\\
D_{ij}(\p,\q) & = & 
8\pi
\int d \q \left(q_i q_j\right) 
 |V_{\p_1,\q}^{\p}|^2 \, n_q \,
\delta_{{\p - \p_1-\q}} \, {\cal L}({\sigma_{\p}
-\sigma_{{\p_1}}-\sigma_{{\q}}})
d \p_{1}.  
\label{KurtsFavoriteEqn}
\EEA This is a Fokker-Planck diffusion equation describing the
diffusion of wave action in the system dominated by nonlocal in
wavenumber interactions.  Comparison between the Fokker-Planck
equation (\ref{KurtsFavoriteEqn}) obtained here, and the similar
Fokker-Planck equation (obtained using WKB theory (Section
\ref{WKBDiffusion} below) will lead to critical  insights into the spectral energy
transfers in internal wave systems and ultimately to the
parametrization of the energy supply to internal wave breaking
processes.

\section{Wave-Wave interactions in the scale separated limit}\label{ScaleSeparations}
In our previous studies \cite{theory} we have seen that under a scale-invariant
assumption, the integrals in the kinetic equation tend to diverge for
small or large wave numbers or both. Therefore the interactions via
extreme scale separations play an important role in energy exchanges in
internal waves. In this section, we are going to develop a rigorous
formalism based on WKB techniques to study such interactions.

\subsection{The Primitive Equations and  Hamiltonian Structure\label{S1}}
\subsubsection{Reynolds decomposition and Hamiltonian structure}

To study interactions between long and short waves we start at (\ref{Canonical}) and make a Reynolds decomposition in wave amplitude: 
\begin{eqnarray}
\Pi \rightarrow \Pi_{0}+ \Pi+ \pi', \ \ \ \ 
\phi=\Phi+\phi' , \ \ \ \ 
\psi = \Psi +\psi' \ .
\label{Reynolds}
\end{eqnarray}
Here the large amplitude waves are represented with $\Pi, \Phi,
\Psi$ and small amplitude waves are given by $\phi'$, $\pi'$ and
$\psi'$.  Given the potentials $\Phi$ and $\phi$, the corresponding velocities
are $${\cal U} = \nabla \Phi  + \nabla^{\perp} \Psi, \ \ \ u'=\nabla \phi' + \nabla^{\perp} \psi'.$$
To simplify the presentation we will utilize the non-rotating approximation ($f=0$) in which $(\nabla^{\perp} \Psi, \nabla^{\perp} \psi) \rightarrow 0 $.  The case of rotating ocean $f\ne 0$ is presented in Appendix. 

We substitute the Reynolds decomposition (\ref{Reynolds}) into the
equations of motion (\ref{PrimitiveEquations}) and subtract equations
for the large amplitude waves. The result is given by
\begin{eqnarray}
\dot\pi' + \nabla \cdot \left((\Pi_{0}+ \Pi+\pi') \nabla \Phi\right)+
 \nabla\left(\pi'\Phi\right)=0, \nonumber \\
\dot\phi'+\frac{|\nabla \phi'|^2}{2}+\nabla\phi' \cdot \nabla\Phi + 
\frac{g}{\rho_0^2} \int\int d\rho' d\rho^{''} \pi'=0. \nonumber \\
\label{EqPrimeVariablesExternal}     
\end{eqnarray}
In these equations $\Pi$ and $\Phi$ are given time-space dependent functions representing the large amplitude waves.

These equations are also {\it Hamilton's equations}, 
\begin{equation}
 \frac{\partial \pi'}{\partial t}  =  \frac{\delta {\cal H}}{\delta \phi'} \, ,
  \qquad
\frac{\partial \phi'}{\partial t}  = -\frac{\delta {\cal H}}{\delta \pi'} \, ,
  \label{Canonical2}
\end{equation}
with the time-dependent Hamiltonian given by 
\footnote{This Hamiltonian may be obtained by substituting (\ref{Reynolds}) to  (\ref{HTL2}). }
\begin{eqnarray}
 {\cal H} =\frac{1}{2} \int d {\bf{r}} \left(
\left( \Pi_{0}+\Pi+\pi' \right) \,
  \left|\nabla \phi' \right|^2  + 2\pi'\nabla\phi'\cdot\nabla\Phi -
  g \left|\int^{\rho} d\hat{\rho}
  \frac{\pi'}{\hat{\rho}} \right|^2
 \right) \, .
  \label{MainHamiltonian} 
\end{eqnarray}
There are two types of terms here - those that will ultimately describe a sea of interacting small scale waves and those that will describe the influence of large amplitude large scale waves on the small scale waves.  In our previous efforts \citep{LT,LT2,theory,Regional,Ultraviolet} these terms are comingled.  Comparing (\ref{HTL2}) and (\ref{MainHamiltonian}), we see that (\ref{MainHamiltonian}) contains additional terms $\Pi$$ |\nabla \phi' |^2$ and $\pi'\nabla\phi'\cdot\nabla\Phi$ that are explicit representations of what will be scale separated interactions.  The term $\pi'\nabla\phi'\cdot\nabla\Phi$ describes the advection of the small-scale internal field by the given large-scale large amplitude field.  The term $\Pi$$ |\nabla \phi' |^2$ represents a coupling of small scales to large through changes in the stratification by the large scale wave.  The term $\pi' |\nabla \phi' |^2$ will represent interactions local in wavenumber.  

We now express the space-dependent variables $\pi^{\prime}(\r),\Pi(\r),\phi^{\prime}(\r)$ and 
$\Phi(\r)$ in terms of their Fourier images $\pi^{\prime}(\p),\Pi(\p),\phi^{\prime}(\p)$ via
(\ref{Fourier}), make the Boussinesq approximation $\frac{\Pi}{\rho}\simeq\frac{\Pi}{\rho_0}$ and
use $\int d \p e^{i \p\cdot \r} =(2\pi)^3 \delta(\p)$ to obtain
\BEA
\H &=& \H_{\rm linear} +\H_{\rm nonlinear},\nonumber\\ 
\H_{\rm linear} &=& \frac{1}{2}\int d \p \left( \Pi_{0} |{\bf k}|^2 |\phi'_{\p}|^2-
\frac{g}{\rho_0^2}\frac{| \phi'_{\p} |^2}{m^2}\right), \nonumber\\
 \H_{\rm  nonlinear}& = &\H_{\rm  local}+ \H_{\rm  sweeping} + \H_{\rm  density}, \nonumber\\
\EEA
\BEA
 \H_{\rm local} &=&-
  {\bf\frac{1}{2(2\pi)^{\frac{3}{2}}}}
\int d \p_1 d \p_2 d \p_3 \delta(\p_1+\p_2+\p_3)
\K_2 \cdot \K_3 \pi'_{\p_1} \phi'_{\p_2} \phi'_{\p_3}, \nonumber\\
\H_{\rm  sweeping} &=&-
  {\bf\frac{1}{2(2\pi)^{\frac{3}{2}}}}
\int d \p_1 d \p_2 d \p_3 \delta(\p_1+\p_2+\p_3)
2 \K_1\cdot\K_2 \Phi_{\p_1} \phi'_{\p_2} \pi'_{\p_3}, \nonumber\\
\H_{\rm  density} &=&-
{\bf\frac{1}{2(2\pi)^{\frac{3}{2}}}}
\int d \p_1 d \p_2 d \p_3 \delta(\p_1+\p_2+\p_3)
\K_2\cdot\K_3 \Pi_{\p_1} \phi'_{\p_2} \phi'_{\p_3}. \nonumber\\
\label{FourierHam} 
\EEA
The Hamiltonian $ \H_{\rm  nonlinear} $ is the sum of three terms:
\begin{itemize}
\item $ \H_{\rm  local}$
represents small amplitudes interacting with small amplitudes. We will call this term ``local'' interactions in anticipation of making a scale separation between large amplitude large scale and small amplitude small scale waves in sections \ref{SweepSection} and\ref{DensitySection}.
\item $\H_{\rm density} $ is the term that describes the variations of stratification that small amplitude  waves experience due to the compression and rarification of isopycnals associated with the
  large amplitude waves. We will refer to this term as a density term.
\item $\H_{\rm sweeping} $ is the term that describes the advection
  (sweeping) of small amplitude waves by large amplitude waves.  In the future,
  we refer to this term as a sweeping term.
\end{itemize}

The main focus of this manuscript is to investigate how the density
and sweeping terms affect the overall spectral energy density.

\subsubsection{Sweeping Hamiltonian\label{SweepSection}}
Following the traditional wave turbulence approach, we make a transformation
to the wave-action variables that represent wave amplitude and
phase:
\begin{eqnarray}
 \phi'_{\bf p}=\frac{i N \sqrt{\sigma_\p}}{\sqrt{2 g} |{\bf k}|}
\left(a_\p-a^*_{-\p}\right)\, , \ \ \ 
\pi'_\p =\frac{\sqrt{g} |{\bf k}|}{\sqrt{2\sigma_\p}N}
\left(a _\p+ a^*_{-\p}\right)\, . 
\nonumber\\
\label{TransformationToActionVariables2}
\end{eqnarray}
We substitute (\ref{TransformationToActionVariables2}) into  
$\H_{\rm sweeping}$ of (\ref{FourierHam}), and obtain
\begin{eqnarray}
  \H_{\rm  sweeping}
  &=&-
  {\bf\frac{1}{2(2\pi)^{\frac{3}{2}}}}
\int d \p_1 d \p_2 d \p_3 \delta(\p_1+\p_2+\p_3)
      2 \K_1\cdot\K_2 \Phi_{\p_1}
      \frac{i N \sqrt{\sigma_{\p_2}}}{\sqrt{2 g}|{\bf k}_2|}
\frac{\sqrt{g} |{\bf k}_3|}{\sqrt{2\sigma_{\p_3}}N}\nonumber\\
&&\hskip 4cm \times\left(a_{\p_2}-a^*_{-\p_2}\right) 
\left(a _{\p_3}+ a^*_{-\p_3}\right)\, . 
\nonumber\\
   \label{HamiltonianSweeping}
      \end{eqnarray}

The next step is algebraically trivial but conceptually fundamental.  We invoke an extreme scale separated limit in which the two small amplitude waves have similar frequency and {\it horizontal wavenumber magnitude}.  {\it No} condition is required on the vertical wavenumber.  This conditioning retains both the induced diffusion (ID) and Bragg scattering (ES) branches of the resonant manifold, figure \ref{Triads}.  

In this scale separated limit of the internal wave problem, $\sigma_{\p_2} \cong \sigma_{\p_3}$ and $|{\bf k}_2| \cong |{\bf  k}_3|$.  Beyond the obvious algebraic simplifications, upon expanding the brackets we find terms of the type $a_{\p_2}a_{\p_3}$, $a^*_{-\p_2}a^*_{-\p_3}$ and $a_{\p_2}a^*_{-\p_3}$, $a^*_{-\p_2}a_{-\p_3}$. In what follows we neglect the $a_{\p_2}a_{\p_3}$ and $a^*_{-\p_2}a^*_{-\p_3}$ terms and retain the $a_{\p_2}a^*_{-\p_3}$, $a^*_{-\p_2}a_{-\p_3}$ terms, since the former terms are nonresonant, while the latter may be in the resonance for some wave numbers.  The discarded terms lead to a process when one lower-frequency wave decays into two high-frequency waves and thus the frequencies do not sum to zero. Such decay is a nonresonant process, so we can remove these terms at the onset.  

After relabelling subscripts, in which $2\to 1$ and $3\to 2$, we obtain
\begin{eqnarray}
\H_{\rm  sweeping} &=&\int  d \p_1 d \p_2 A_{\rm  sweeping}(\p_1, \p_2) a_{\p_1}a^*_{\p_2}, 
\nonumber\\
&{\rm with }& \\
A_{\rm  sweeping}(\p_1, \p_2)  
&  =& -\frac{1}{2}i 
\left({\K_1-\K_2}\right)\cdot\left( \K_1+\K_2\right) \Phi_{\p_1-\p_2} \; . \nonumber
\label{KernelSweep}
\end{eqnarray}
In the limit of large vertical background scales, (\ref{KernelSweep}) describes a quasi-coherent translation of small scale small amplitude waves by the large scale background.  At $1/2$ the vertical scale of $\p_1$ and $\p_2$, (\ref{KernelSweep}) describes a Bragg Scattering process.  This is distinct from 'local' interactions as the large amplitude wave has a much larger horizontal scale.

\subsubsection{Density Hamiltonian}\label{DensitySection}
We now can repeat the same steps for the density Hamiltonian.  We
substitute (\ref{TransformationToActionVariables2}) into the $\H_{\rm
  density} $ of (\ref{FourierHam}), use
$\sigma_{\p_2} \cong \sigma_{\p_3}$ and $|{\bf k}_2| \cong |{\bf k}_3|$,
relable subscripts and obtain
\begin{eqnarray}\label{KernelDensity}
\H_{\rm density} &=&\int d \p_1 d \p_2 A_{\rm density}(\p_1, \p_2) a_{\p_1}a^*_{\p_2}, 
\nonumber\\ 
&{\rm with }& \\
A_{\rm density}(\p_1, \p_2) & = & \frac{1}{2 \Pi_{0}}  \Pi_{\p_2-\p_1} \sqrt{  \sigma_{\p_1}\sigma_{\p_2}} \; . 
\nonumber 
\end{eqnarray}

In the limit of large vertical background scales, (\ref{KernelDensity}) describes the modulation of the background stratification.  At $1/2$ the vertical scale of $\p_1$ and $\p_2$, (\ref{KernelDensity}) describes a Bragg Scattering process.  This is distinct from 'local' interactions as the large amplitude wave has a much larger horizontal scale.

\subsubsection{Quadratic Hamiltonian for inhomogeneous wave turbulence}

Let us neglect the local interaction term $ \H_{\rm local}$ in
Hamiltonian (\ref{FourierHam}). Then the Hamiltonian
(\ref{FourierHam}) of small amplitude small horizontal scale internal waves $a_\p$ superimposed
into a field of large amplitude large horizontal scale internal waves given by
space-time-dependent $\Pi$ and $\Phi$ are given by a form that is
quadratic in the small-scale variables $a_{\p}$
\begin{eqnarray}
\H &=&\int  d \p_1 d \p_2 A(\p_1, \p_2) a_{\p_1}a^*_{\p_2}, \nonumber 
\label{HamiltonianGeneral}\\
&{\rm with }& \\
 A(\p_1,\p_2)&=& \OMEGAP \delta(\p_1-\p_2) + A_{\rm  sweeping}(\p_1, \p_2)+ 
A_{\rm density }(\p_1, \p_2), \nonumber 
\label{ResultingHamiltonian}
\end{eqnarray}
where  $ A_{\rm sweeping}(\p_1, \p_2)$ and $A_{\rm density }(\p_1, \p_2)$ are given in (\ref{KernelSweep}) and (\ref{KernelDensity}).  
The $A(\p_1,\p_2)$ 
are time dependent and depend upon the phases of the external field. 

\subsection{WKB approach\label{S2}}
At this point, we have a three-wave Hamiltonian (\ref{ResultingHamiltonian}) with one large-amplitude large horizontal scale wave interacting with two smaller amplitude smaller horizontal scale waves that have similar frequencies.  Below we assume a nearly-resonant paradigm and perform algebraic manipulations that take into account the induced diffusion portion of the resonant manifold.  

\subsubsection{The Wave Packet Transport Equation}\label{sec:PacketTransport}
In the spatially homogeneous wave turbulence of Section \ref{WTSection} we used wave action density 
$n_\p$ and a linear dispersion relation $\sigma_\p$, both being a
function of a wave number $\p$. In the case when there is a slowly
varying large scale background, i.e. a system where spatial
inhomogeneity is present, the properties of wave action and the dispersion
relation will depend on the position in space. Then it makes sense to
introduce an additional parameter, a position vector $\r$ in wave action
and linear dispersion relation.  The theory for this spatial dependence is developed in \cite{WKBHamiltonian} using a Gabor transform to represent the envelope structure describing the spatial localization and carrier frequency.  The leading order balance in \cite{WKBHamiltonian} leads to action and phase conservation along ray paths.  The balance on the envelope scale implies phase modulation, for which we direct the reader to the appendix of \cite{CohenLee} for clarity.  Associated with the envelope structure is a residual circulation \cite{BM05}.  The potential for the wave packet to interact with its envelope structure is possible \cite{BM05,dosser2011anelastic} but would require a modification of the uniform potential vorticity statement (\ref{PV}).  It is at this stage that one might also want to consider the potential for nonlinear wave steepening effects associated with $\H_{\rm local}$ to counter the dispersion of ray characteristics as a precursor to the description of solitary wave dynamics.  

The familiar statement of action conservation that we are after is obtained in \cite{WKBHamiltonian} by assuming the scale of the envelope structure is large in comparison to the inverse wavenumber of the small scale wave, i.e. the wave packet contains many oscillations.  The result required here can be obtained more simply by using a Wigner transform alone to define the space-time dependent wave action spectral density:  

\BEA\label{Wigner}
n_{\p,\r}\equiv
\int e^{i\q\cdot\r} \langle a_{\p+\q/2} a_{ \p-\q/2}^*\rangle d \q \; ,
\EEA
in which the transform variable $\q$ is a difference between two large wave numbers $\p_1$ and $\p_2$, $\q=\p_1-\p_2$ and the field variables $a$ are evaluated at $\p \pm \q/2$.  
We similarly introduce the space-dependent intrinsic frequency $\omega_{\p,\r}$ 
\BEA
\omega_{\p,\r}=\int e^{ i \r\cdot\q}A({\p+\q/2,\p-\q/2}) d\q,\label{wkx}
\EEA
This is a generalization of our ``traditional'' wave action (\ref{WaveAction}) which allows for slow variations of the background.  It will incorporate the large vertical scale contributions of ${\cal H}_{{\rm sweeping}}$ and ${\cal H}_{{\rm density}}$.  

To derive the transport equation, we take definition
(\ref{Wigner}), differentiate it with respect to time, use Hamilton's
equation of motion (\ref{fieldequation}), with the Hamiltonian
(\ref{HamiltonianGeneral}).  The spatial dependence is then made explicit by representing $A(\p_1,\p_2)$ and $a(\p_1)a^{\ast}(\p_2)$ with their corresponding Fourier transforms.  After an inspired change of variables \citep{lvov1977fundamentals,WKBHamiltonian} that maps the ID portion of the resonant manifold, one obtains an intermediary expression:
\begin{equation}\label{Intermediary}
i \frac{\partial n_{\p,\r}}{\partial t} = \int \frac{d\r^{\prime}d\r^{\prime\prime}}{(2\pi)^6} dq^{\prime}dq^{\prime\prime} e^{iq^{\prime}\cdot(\r-\r^{\prime}) + iq^{\prime\prime}\cdot(\r-\r^{\prime\prime}) } 
\LSBA 
\omega_{\p+\frac{q^{\prime\prime}}{2}}(\r^{\prime}) n_{\p-\frac{q^{\prime}}{2}} (\r^{\prime\prime})  - 
\omega_{\p-\frac{q^{\prime\prime}}{2}}(\r^{\prime}) n_{\p+\frac{q^{\prime}}{2}}(\r^{\prime\prime})
\RSBA \end{equation}
After expanding $\omega$ and $n$ in Taylor series with respect to $\p$ and truncating higher order terms, one
ultimately arrive at the action balance:
\BEA
\frac{\partial n_{\p,\r}}{\partial t}+
\nabla_\p \sigma_{\p,\r}\cdot 
\nabla_{\r}n_{\p,\r}-\nabla_\r\sigma_{\p,\r}\cdot\nabla_{\p}n_{\p,\r}=0,\nonumber\\ %
\label{TransportEquation}
\EEA
or alternately
\BEA
\frac{\partial n_{\p,\r}}{\partial t}+
\nabla_{\r} \cdot \LSBA n_{\p,\r} \nabla_{\p} \sigma_{\p,\r} \RSBA  -  \nabla_{\p} \cdot \LSBA n_{\p,\r} \nabla_{\r} \sigma_{\p,\r} \RSBA = 0.\nonumber\\ %
\label{TransportEquationDu}
\EEA 
Integration of (\ref{TransportEquationDu}) over wavenumber provides a connection to the space-time variational formulations found in \cite{Witham}, Chapters 11.7 and 14.  The Bragg scattering process residing in the scale-separated Hamiltonian (\ref{KernelSweep}) has been eliminated by using the Wigner transform and Taylor series expansion that serendipitously exploits the symmetries associated with the ID resonance.  

\subsubsection{Application for internal waves}\label{sec:ApplicationIW}
We now take the expression for $A_{\p_1, \p_2}$ from 
(\ref{HamiltonianGeneral}) and substitute it into 
(\ref{wkx}) for the space-dependent linear dispersion relation. 
The result is given by:  
\BEA
\omega_{\p,\r}= \OMEGAP - \K\cdot {\cal U}(\r) + \OMEGAP\frac{\Pi(\r)}{2\Pi_0},
\label{SpaceDependentFrequency} 
\EEA
where $\p=({\bf k},m)$ is the wavevector, ${\cal U}(\r)$ is the time-dependent horizontal velocity
of the external large-scale wavefield and $\Pi(\r)$ is the time dependent stratification
of the external wavefield. Here the $\OMEGAP$ term was produced by the
term proportional to $\delta(\p_1-\p_2)$, $\K\cdot{\cal U}$ comes from the sweeping term
in the Hamiltonian and $\OMEGAP \Pi / 2 \Pi_0 $ comes from density term in the Hamiltonian.  The frequency $\OMEGAP$ is given by (\ref{InternalWavesDispersion}) using $f=0$.  For a high-frequency wave in a rotating ocean, $\cal{U}$ is replaced by $\nabla \Phi + \nabla^{\perp} \Psi$.  The derivation for $f \ne 0$ appears in the Appendix.  


\subsubsection{The Ray Path Wave Packet Transport Equation}\label{RayTracing}
The transport equation (\ref{TransportEquation}) or its alternative
formulation (\ref{TransportEquationDu}) with the space-time dependent
linear dispersion relationship (\ref{SpaceDependentFrequency}) is a
fundamental result expressing wave action conservation that provides
the basis for the analyses to follow.

The action balance (\ref{TransportEquation}) can be simply solved by the method of
characteristics.  The characteristics of (\ref{TransportEquation}),
also called rays, are defined by
\BEA
\dot \r(t) & \equiv & \nabla_\p \sigma_{\p,\r} ,
\ \ 
\dot \p(t)  \equiv  - \nabla_\r\sigma_{\p,\r} \; . 
\label{Characteristics}
\EEA  
Equations (\ref{Characteristics}) imply that wave action spectral density is conserved along these characteristics:
\BEA 
\frac{\partial n_{\p,\r}}{\partial t}+
\dot \r \cdot \nabla_{\r}n_{\p,\r}+\dot\p \cdot \nabla_{\p}n_{\p,\r}=0.  
\label{TransportEquationChar}
\EEA This is the classical representation for the conservation of
action spectral density along ray trajectories.  A derivation for a
general Hamiltonian set is presented in \cite{WKBHamiltonian}, here the
derivation is specifically for internal waves in the scale separated
limit.  Integration over wavenumber provides the space-time result
presented in \cite{Witham}.  This result often appears as an analogy
to Louiville's theorem for the conservation of the phase volume of
particles without justification.  Note that it is the balance of first-order terms in a
Taylor series expansion of (\ref{Intermediary}).

\subsubsection{An Ensemble Path Transport Equation}\label{WKBDiffusion}

In what follows we derive a combined advective-diffusive transport equation for the
action balance (\ref{TransportEquation}) by generalizing an approach
found in \cite{SnazarNonlocal1}.  There are strong parallels here to
the discussion of \cite{Taylor} that appear in the appendix of \cite{MB77}: The illuminating analogy with particle dispersion in \cite{Taylor} is to substitute wavenumber $\p$ for the Lagrangian particle position $\r$ and obtain a quantitative approach for discussing the migration and dispersion of wave packets
in the spectral domain following ray trajectories.  Having said this, it should be intuitively obvious that, since particle dispersion in $\r$ has little to do with resonance, neither should the issue of dispersion of $\p$ in phase space be intrinsically tied to resonance!  Yet, an emphasis on the resonant paradigm is the interpretive context pursued in \citep{MB77, SnazarNonlocal1}.  A second key departure is that our motivation stems from the fact that the GM76 spectrum is a no-flux solution to the Fokker-Planck equation (\ref{KurtsFavoriteEqn}).  We acknowledge the oceanographic literature in this regard and so are focused upon the issue of a mean drift of wave packets to high wavenumber that is accessible in ray-tracing simulations \citep[e.g.][]{HWF} but not explicitly represented in (\ref{KurtsFavoriteEqn}).  To underscore the distinction, the existence of a mean drift implies {\em relative} dispersion (e.g. \cite{bennett1984relative}) rather
than the issue of {\em absolute} dispersion addressed in \cite{Taylor}.  

The very first step in our derivation is to specify a wave-packet ensemble mean drift and departures from the mean drift.  This decomposition is an improvement on the arguments presented in \cite{MB77} and \cite{SnazarNonlocal1}.  Our limited knowledge of the ray literature does not permit us from commenting on the originality of our interpretation.  It is, however, crucial in understanding intrinsic  differences between the kinetic equation and ray theory.  Given the nature of our understanding, the issue assuredly carries over to other physical problems such as the interaction of near-inertial waves with lower frequency flows, \cite[e.g.][]{YBJ, kafiabad2019diffusion, Dong2020frequency} surface gravity wave interaction with lower frequency flows \citep{boas2020directional} and Rossby wave - Rossby wave interactions \cite{nazarenko2011wave}.  

We represent the wave action of a single wave packet as a sum of a spatially homogeneous part $\overline{n}$ and small ``wiggles'' $\tilde{n}$:
\BEA
n_{\p,\r} = \overline{n}_\p + \tilde{n}_{\p,\r};\ \ \  n_\p=\int d\r \; n_{\p,\r}; \ \ \ \int d\r \; \tilde{n}_{\p,\r} =0 , 
\label{wiggles}
\EEA
and acknowledge the presence of a mean drift in the spectral domain by adding zero:
\BEA
\dot{\p} = \dot{\p} - \langle \dot{\p} \rangle  + \langle \dot{\p} \rangle \; ,
\label{AddZero}
\EEA in which $\langle \dots \rangle$ is an ensemble average for a
system with spatially homogenous statistics, so that $\langle \dots
\rangle$ is independent of $\r$.  The mean drift arises due to inhomogeneities of the ray path statistics in the {\em spectral} domain.  Please note that the dimensions of
$\overline{n}_\p$ and $n_{\p}$ are different.

Starting from the flux form of the action balance (\ref{TransportEquationDu}), we substitute (\ref{wiggles}), (\ref{AddZero}), invoke an ensemble average and integrate over $\r$ to obtain
\BEA
\frac{\partial \langle n_{\p} \rangle }{\partial t}=
- \int d\r  \langle \nabla_\p\cdot[  \dot{\p} - \langle \dot{\p} \rangle ] n_{\p,\r} \rangle - \nabla_\p \cdot [ \langle \dot{\p} \rangle \langle n_{\p} \rangle ] .  
\label{TransportEquation3}
\EEA

Closure of this equation depends upon writing $n_{\p,\r}$ in terms of $n_{\p}$.  At this juncture, we invoke that property that wave action spectral density does not change
along trajectories:
\BEA 
n_{\p,\r} \equiv n(\p(t),\r(t),t) = n(\p(t-T),\r(t-T),t-T)  \nonumber\CR
= \overline{n}\Big[\p(t) - \int\limits_{t-T}^t \dot p(t') d t'; \; t-T\Big] 
+ \tilde{n}\Big[\p(t) - \int\limits_{t-T}^t \dot p(t') d t'; \; \r(t) - \int\limits_{t-\tau}^t \dot \r(t') d t'; \;t-T\Big] \nonumber\CR
\EEA
We then execute a Taylor series expansion of $\overline{n}$, 
\BEA
n(\p(t-\tau),\r(t-\tau),t-\tau) \simeq 
\overline{n}(\p(t); \; t-T) - \nabla_p \overline{n}(\p(t); \; t-T) \cdot \int\limits_{t-T}^t \dot{ \p}(t') d t'+ \tilde{n}(\dots) \; ,
\nonumber\CR
\label{NazarenkoExpansion}\EEA
substitute, add zero once again, and using the definition of ensemble averaging $\langle \dots \rangle$
\BEA
\Big\langle \nabla_{\p} [ \dot{\p} - \big\langle \dot{\p} \big\rangle] \overline{n} (\p(t), t-\tau) \Big\rangle & \cong & \nabla_{\p} \big\langle \dot{\p} - \big\langle \dot{\p} \big\rangle \big\rangle \big\langle \overline{n} (\p(t), t-\tau) \rangle =  0 \nonumber
\EEA
and
\BEA
\Big\langle \nabla_{\p} [ \dot{\p} - \big\langle \dot{\p} \big\rangle] \int\limits_{t-\tau}^t \big\langle \dot{\p} \big\rangle dt' \cdot \nabla_{\p}  \overline{n} (\p(t), t-\tau) \Big\rangle & \cong & 
\nabla_{\p} \big\langle  \dot{\p} - \big\langle \dot{\p} \big\rangle \big\rangle \int\limits_{t-\tau}^t \big\langle \dot{\p} \big\rangle dt' \cdot \nabla_{\p}  \big\langle \overline{n} (\p(t), t-\tau) \big\rangle \nonumber\CR
 & = & 0
\EEA
as $\langle \dot{\p} - \langle \dot{\p} \rangle \rangle \equiv 0$ and neglect an initial transient term
\BEA
\langle \nabla_{\p} [ \dot{\p}(t) - \langle \dot{\p} \rangle] \tilde{n}(\p(t-\tau),\r(t-\tau),t-\tau) \rangle
\EEA
so that (\ref{TransportEquation3}) becomes 
\BEA
\frac{\partial \langle n_{\p} \rangle}{\partial t}=
- \nabla_\p \cdot \int\limits_{t-\tau}^t \langle [ \dot{\p}(t) - \langle \dot{\p} \rangle] [  \dot{\p}(t'-\tau) - \langle \dot{\p} \rangle ]\rangle dt' \nabla_\p \langle n_{\p} \rangle - \nabla_\p \langle \dot{\p} \rangle \langle n_{\p} \rangle,
\label{FokkerPlanckImproved}
\EEA
We introduce the auto-lag covariance matrix \BEA\label{Autolag} {\cal C}_{ij}(\p,t,t') = \Big\langle
\Big[ \dot \p(\r(t)) - \langle \dot \p(\r(t)) \rangle\Big]_i \Big[\dot \p(\r(t'))
-\langle \dot \p(\r(t'))\rangle \Big]_j\Big\rangle.  \EEA

The final result is then \BEA \frac{\partial
  \langle n_{\p} \rangle}{\partial t}= - \nabla_{\p_i}\cdot \int\limits^t_{t-\tau} {\cal C}_{ij}(\p,t,t^{\prime}) dt^{\prime} \cdot
\nabla_{\p_j} \langle n_\p \rangle \;\; - \;\; \nabla_{\p_i} \langle \dot \p_i \rangle \langle n_\p \rangle.
\label{FokkerPlanckFINAL}
\EEA

Convergence of the time integral, in which one can replace the lower limit of integration by $-\infty$, is the hallmark of a Markov approximation which we investigate further in \cite{polzinCompanion}.  This equation encapsulates our fundamental theoretical result: The transport equation changes from diffusion (\ref{KurtsFavoriteEqn}) to an expression that involves both advection and diffusion in the spectral domain.  Instead of being a no-flux stationary state, the GM76 spectrum, for which $\langle n_\p \rangle \propto m^0$, now supports a downscale action flux.  

Our decomposition of wavenumber tendency into mean drift and dispersion about that mean drift provide a concrete mathematical interpretation for an interaction timescale $\tau_i$:
\begin{equation}\label{TauI}
    \tau_i^{-1} = | \langle \dot{\p} \rangle | / | \p |
\end{equation}
and correlation time scale $\tau_c$:
\begin{equation}\label{TauC}
    \tau_c = \int\limits^t_{-\infty} {\cal C}_{ii}(\p,t,t^{\prime}) dt^{\prime} / \langle | \dot \p(\r(t)) - \langle \dot \p(\r(t)) |^2 \rangle  \rangle . 
\end{equation}
Intuitive notions of the interplay between $\tau_i$ and $\tau_c$ are discussed in \cite{M86} and in \cite{SnazarNonlocal1} using expressions for ${\cal C}_{ij}$ (\ref{Autolag}) in which the ensemble mean drift has not been subtracted.  
The sentiment in \cite{M86} is that a separation between $\tau_c$ and $\tau_i$ is problematic for the oceanic internal wavefield.  Numerical ray tracing results \citep{HP83, HPM} do not elucidate why this might be.  Our decomposition of wavenumber tendency into mean drift and dispersion about that mean drift, the revised Fokker-Planck (\ref{FokkerPlanckFINAL}) and revised covariance matrix (\ref{Autolag}) provide a concrete mathematical interpretation for such judgements about $\tau_i$ and $\tau_c$. 

Having summoned the analogy between particle dispersion \cite{Taylor} and dispersion of wavepackets in wavenumber, we are led to a degree of skepticism concerning \cite{MB77}'s and \cite{SnazarNonlocal1}'s interpretation that ray tracing should collapse onto the Fokker-Planck and diffusivity derived from the resonant kinetic equation (\ref{KurtsFavoriteEqn}).  Convergence of the time-lagged auto-covariance (\ref{TauC}) relates a finite diffusivity to the product of a covariance and correlation time scale.  If one casts this as a resonant process, the covariance will be infinitely small and the correlation time scale infinitely long, thus leading to an inconsistency between interaction and correlation time scales in the resonant limit.  Introducing a broadened kinetic equation (\ref{DIA}) with finite bandwidth helps so much as it changes the ratio of correlation time scale to interaction time scale from infinity to something large, but is not a resolution.  As one runs to the finite amplitude of the weakly nonlinear problem, the resonant bandwidth becomes the rms Doppler shift \citep{Ultraviolet}.  This is aphysical.

We find through numerical experimentation in \cite{polzinCompanion} that the mean drift $\langle \dot{\p} \rangle$ is a resonant process and dispersion about the mean drift ${\cal C}_{ij}$ is non-resonant.  The latter should not come as a surprise once one appreciates the direct analogy between particle dispersion and ray tracing originally suggested in \cite{MB77}:  particle dispersion in turbulence has nothing to do with the concept of resonance.  The moments $\langle \dot{\p} \rangle$ and ${\cal C}_{ij}$, and changes in the structure and scaling of the resonant bandwidth, are the signature differences of ray theory vs the kinetic equation, paralleling differences between Eulerian \citep{kraichnan1959structure} and Lagrangian \citep{kraichnan1965lagrangian} representations of 3-D turbulence.  These differences are rooted in the distinctions between amplitude-modulated and frequency-modulated signals.

\section{Energy Transport in oceanic internal waves}\label{sec:EnergyTransport}
In \cite{Regional,Ultraviolet} we note the tension between an apparent pattern match between observed spectral power laws being in apparent agreement with stationary states of the Fokker-Planck equation derived
from the kinetic equation (\ref{BroadenedKineticEquation}):
\begin{eqnarray}
\frac{\partial n(\p)}{\partial_m} & + & \frac{\partial}{\partial_m} D_{33} \frac{\partial}{\partial_m} n(\p) = 0 
\label{eq:OneDim}
\end{eqnarray}
and this result being inconsistent with what is observationally understood about the energy sources and sinks.  Those stationary states come from asserting a balance only in vertical wavenumber, for which there are two families: no-flux states for which $n(\p) \propto k^{-x}m^{-y}$ with $y=0$ and constant flux states for which a linear relationship between $x$ and $y$ attains.  Both families are oceanographically relevant \cite{Regional} and, more to the point, the Garrett and Munk model (GM76) is a member of the no-flux family.  This no-flux result attains simply because that spectrum ($(x,y)=(4,0)$) has no gradients in action in vertical wavenumber.

The ray path perspective moves away from this interpretation so that downscale transport is closed as an advective transport (\ref{FokkerPlanckImproved}).  Below we explore the downscale energy transport of this advective contribution for the GM76 spectrum, for which
\begin{equation}
       D_{33} = \frac{2}{\pi} \frac{k m^2 e_0 m_{\ast}}{N} \; . 
\end{equation}
Here $k$ is horizontal wavenumber magnitude, $e_0=0.0030$ m$^2$ s$^{-2}$ is the total energy, $m_{\ast} = 4 \pi / 1300$ m$^{-1}$ is a bandwidth parameter, $N = 0.0052$ s$^{-1}$ is buoyancy frequency and the underlying energy spectrum $e(m,\sigma)$ appears, for example, as equation (5) in \cite{Ultraviolet}.  {\em If} we associate the mean drift $\langle \dot{m} \rangle$ with $\partial_m D_{33}$ derived from the kinetic equation (\ref{KurtsFavoriteEqn}) \citep{polzinCompanion} and include a
factor of two to account for the two-sided spectral representation, the downscale energy transport is
\begin{equation}
\mathcal{P} = 2 \int_f^N \langle \dot{m} \rangle  e(m,\sigma) d\sigma = 2 \big( \frac{2}{\pi} \big)^2 \big( \frac{e_0 m_{\ast}}{N} \big)^2 f \log(\frac{N}{f})\cong 1.0\times10^{-8} [{\rm W \; kg}^{-1}]
\label{Power}
\end{equation}
which, apart from the prefactor of $1.0\times10^{-8}$ being an order of magnitude too large, is virtually identical to the finescale parameterization \cite{Finescale}, their equations 27 and 40.  In \cite{polzinCompanion} we demonstrate through a path integral closure that, indeed, $\langle \dot{m} \rangle = \partial_m D$ for the one-dimensional representation of high frequency internal waves interacting with inertial waves.  We further demonstrate that both are consistent with simple scale invariant ray-tracing numerical simulations that treat the interaction as a one-dimensional problem.  We believe that this one-dimensional treatment is a reasonable representation of extreme scale separated interactions.  The one-dimensional version of (\ref{KurtsFavoriteEqn}), (\ref{eq:OneDim}), dates to the dawn of modern oceanography and is supported by basic scale analysis \citep{MB77, SK99a}.  It is underpinned by the integrable singularity of the inertial peak in the internal wave frequency spectrum and the lack of horizontal velocity gradients in that peak that is encoded in the dispersion relation.  A modern analysis of local and extreme scale separated interactions in a non-rotating context \citep{dematteis2022origins} assigns an energy transport associated with horizontal and diagonal terms of the diffusivity tensor that are an order of magnitude smaller than the Finescale Parameterization, two orders of magnitude smaller than (\ref{Power}).  These non-vertical transports are likely overestimates in a rotating paradigm due to the vanishing of horizontal velocity gradients at the inertial peak.  

The result (\ref{Power}) stands in dramatic contrast with numerical results concerning the more general problem of high-frequency internal wave packets refracting in a background sea of internal waves \citep[e.g.][]{HWF,SK99b,ijichi2017eikonal}.  This contention has required a systematic examination and physical interpretation of the assumptions within both kinetic equations (Section \ref{WTSection}) and ray-path (Section \ref{S2}) approaches to pinpoint the multiple junctures which might underpin a systematic difference between observation and theory concerning extreme scale separated interactions that are presented above in (\ref{Power}).

To date we have identified three potential soft spots that could resolve the contradiction. 

The first is that {\it both} the ray tracing and kinetic equation discard a coupling between leading order processes that leads to a subtractive cancellation of these leading orders.  These leading orders are provided by the induced diffusion mechanism and the Bragg scattering mechanism, in which the phase locking introduced by induced diffusion is damped by Bragg scattering.  Both are part of an extreme scale separated Hamiltonian (\ref{KernelSweep}); Bragg scattering is discarded in a Taylor series expansion about the induced diffusion resonance (\ref{Intermediary}) that produces the action conservation statement {\ref{TransportEquation}).  Representation of this damping process can be defined by bundling eight distinct triads from the resonant manifold; the standard three wave kinetic equation is a perturbation expansion in wave amplitude limited to one triad through a 'random phase' approximation to obtain (\ref{BroadenedKineticEquation}).  
This coupling is pursued in the companion manuscript \cite{polzinCompanion}, where we argue that these {\it new} physics are the route through which the ocean resolves the contradiction.  

The second is that there is the potential for finite amplitude effects in which interactions, both resonant and non-resonant, represent a stochastic forcing in phase space on a short time scale that disrupts the phase-velocity - group velocity resonance on a long time scale.  This requires assessment of the resonant bandwidth, which  differs in Eulerian (kinetic equation, (\ref{ResonanceWidth})) versus ray-coordinates, in combination with the amplitude and decorrelation time scales of that forcing.  This is investigated more fully in \cite{polzinCompanion}.  Our opinion is that (\ref{Power}) and associated scaling is a fundamental metric that should be recoverable by kitchen sink efforts as a small amplitude limit of wave turbulence and using a scale separation that aligns with the assumptions underpinning ray tracing. We offer the opinion that this is how the kitchen sink numerics of ray tracing resolves the contradiction.  

The third is that extant efforts at ray tracing high-frequency internal wave packets refracting in a background sea of internal waves (\cite{HWF,SK99b,ijichi2017eikonal}) can not be relied upon as a robust arbiter of this contradiction.  Ray tracing is an asymptotic method requiring a scale separation in horizontal wavenumber (\ref{SweepSection}) in addition to the spatial averaging implied in the envelope structure (\ref{sec:PacketTransport} \cite{WKBHamiltonian}).  The ray tracing numerical studies acknowledge none of this and regard scale separation as a tunable parameter.  They consistently document sensitivity to the specification of the scale separation and consistently find that the observed finescale metric of energy sourced to turbulent dissipation (\cite{Finescale}) requires a scale {\it equivalence}, i.e. requires the small parameter of an asymptotic expansion to be $ \sim O(1)$.  This is the hallmark of interactions represented as $\H_{\rm local}$ (\ref{FourierHam}) that are spectrally local in wavenumber and need to be treated by other methods \cite{dematteis2021downscale,dematteis2022origins}.

\section{Conclusions}\label{sec:Conclusions}
We have presented two distinct derivations of transport equations for the refraction of high frequency internal waves in inertial wave shear.  One derivation results from 'standard' wave turbulence techniques with the addition of near-resonant interactions and describes the wavefield as a system of amplitude modulated waves.  This kinetic equation-based derivation results in a Fokker-Plank equation which returns an estimate of no-net downscale transport in vertical wavenumber for the canonical spectrum of oceanic internal waves referred to as the Garrett-Munk spectrum and small transports associated with horizontal and off-diagonal elements \citep{dematteis2022origins}.  The second derivation is based on ray-tracing techniques in the WKB limit.  Here the ensemble-averaged transport equation (\ref{FokkerPlanckFINAL}) contains a mean drift term that is absent from the Fokker Plank equation derived from the kinetic equation (\ref{KurtsFavoriteEqn}).  This term leads to a prediction of ocean dissipation (\ref{Power}) an order of magnitude greater than supported by ocean observations.

The disparity of these results can be interpreted from an analogy between ray characteristics and Langrangian paths.  Recognizing this distinction improves earlier derivations of the ray path transport equation in \cite{MB77} and \cite{SnazarNonlocal1}:  it provides a rigorous basis for the intuitive characterization of the transport of energy to dissipation invoked in \cite{HWF,SK99b,ijichi2017eikonal}.  We arrived at this rigorous result by adding zero and explicitly invoking an ensemble average.

\cite{Holloway1,Holloway2} argue that internal wave interactions might not be sufficiently weak for wave turbulence theory to be valid.  That commentary is directed at inferences the decay rate of narrowband perturbations to the spectrum being much larger than the internal wave frequency \cite{M86}.  This is inconsistent with the express intent that the kinetic equation describes the slow evolution of the wave spectrum.  In \cite{Ultraviolet} we identify this decay rate is as the small amplitude limit of the resonant bandwidth $\Gamma$ (\ref{ResonanceWidth}).  However, the differences in our two Fokker-Planck equations do not hinge upon this issue.  We have demonstrated in \cite{Ultraviolet} that, at finite amplitude, the bandwidth is proportional to the rms Doppler shift.  This denotes a degenerate state in which the bandwidth describes the quasi-coherent translation of small-scale waves, i.e. sweeping', rather than their interaction.  In the WKB-based derivation, we operate upon the Hamiltonian with a Wigner transform that integrates over that narrow band perturbation and its nearly resonant decay partners in an extreme scale-separated limit.  This removes the apparent discrepancy and arrives at different notions of bandwidth and the role of off-resonant interactions \cite{Ultraviolet}.  We investigate these issues in greater detail in \cite{polzinCompanion}.

As we look back over the landscape of this endeavor, what we have is a well-established metric for ocean mixing known as the Finescale Parameterization \citep{Finescale}.  At best, the Finescale Parameterization is underpinned by a heuristic description as an advective spectral closure \citep{P04a} in the context of an energy transport equation that eschews action conservation in which energy transport in horizontal wavenumber keeps pace with that in vertical wavenumber.  This interpretation contrasts with the pivotal role that induced diffusion was perceived to play in determining downscale transports in vertical wavenumber only.  A possible resolution can be found in recent characterizations of the internal wave kinetic equation \citep{dematteis2021downscale,dematteis2022origins} that coincide in magnitude and scaling with the description of transports encapsulated within the Finescale Parameterization.  That work emphasizes the importance of local interactions.  

Here we have derived a transport equation (\ref{FokkerPlanckImproved}) based upon a packet ensemble that contains an advective transport term.  When the advective transport term is evaluated in a rotating context, it gives rise to a transport estimate (\ref{Power}) that is an order of magnitude greater than the Finescale Parameterization.  This inconsistency will be analyzed in \cite{polzinCompanion}, where we suggest that the fourth-order cumulants, which are subleading to the product of two two-point correlators inhomogeneous wave turbulence, are playing a significant role in the spatially inhomogeneous ray-coordinate system.

\newpage

\bibliographystyle{jfm}

 \section{Appendix: Rotations Included\label{RotationsIncluded}}

We now repeat all the calculations with rotations included. We start
from the primitive equations (\ref{PrimitiveEquations}), decompose the velocity using (\ref{decomp}), and then use the expression for potential vorticity (\ref{PotentialVorticity}) to obtain:  
\begin{eqnarray}  
 \Pi_t + \nabla \cdot \left(\Pi\, \left(\nabla \phi + \nabla^{{\perp}} \Delta^{-1}
 \left(\frac{f}{\Pi_{0}} \Pi-f\right)\right)\right)
  &=& 0 \, , \nonumber\\
  \phi_t + \frac{1}{2} 
  \left|\nabla\phi+ \nabla^{{\perp}}\Delta^{-1} \left(\frac{f}{\Pi_{0}} \Pi-f\right)
  \right|^2 
  && \nonumber \\ +
   \Delta^{-1} \nabla \cdot
   \left[ \frac{f}{\Pi_{0}} \Pi \,
    \left(\nabla^{{\perp}}\phi - 
    \nabla\Delta^{-1}\left(\frac{f}{\Pi_{0}} \Pi-f\right)\right) \right] 
    && \nonumber \\    + \frac{g}{\rho}
   \int^{\rho}\int^{\rho_2} \frac{\left(\Pi-\Pi_{0}\right)}
       {\rho_1} \, d\rho_1 \,
    d\rho_2  &=& 0 \, .\nonumber\\ \label{PrimitiveEquationHAM}
\end{eqnarray}
We now substitute the Reynolds decomposition (\ref{Reynolds}) into  (\ref{PrimitiveEquationHAM}). In doing so we use the fact that potential 
vorticity is assumed to be conserved on isopycnals: 
$$\frac{f}{\Pi_{0}} = \frac{f+\Delta\Psi +\Delta\psi'}{\Pi_{0}+ {\Pi} + \pi'} \; .$$ 

Here we
denote by $\Psi +\psi'$ the divergence-free part of the velocity
field: $$\Psi + \psi' = \Delta^{-1}\left(\frac{f}{\Pi_{0}}(\Pi_{0}+{ \Pi}+\pi')\right).$$
Then
\begin{eqnarray}  
 \dot\pi'&+& \nabla \cdot \left( (\Pi_{0} + \Pi + \pi') 
\left(\nabla \phi' + \nabla^{\perp} (\Delta^{-1}(\frac{f \pi'}{\Pi_{0}} )) \right)\right)+ 
\nabla\cdot\left(\pi'\left(\nabla\Phi+\nabla^\perp(\Delta^{-1}(\frac{f\Pi)}{\Pi_{0}})\right)\right)
  = 0 \, , \nonumber\\
  \dot\phi' &+& \frac{1}{2} 
  \left|\nabla\phi'+ \nabla^{\perp} (\Delta^{-1}(\frac{f \pi'}{\Pi_{0}} )  \right|^2
+\Delta^{-1}\nabla\cdot
            \left(\frac{f}{\Pi_{0}} (\Pi_{0}+\Pi+\pi')(\nabla^\perp\phi'-\nabla (\Delta^{-1}(\frac{f \pi'}{\Pi_{0}} ))\right) 
 \nonumber\\
&&+\Delta^{-1}\cdot\ \nabla\left( \frac{f \pi'}{\Pi_{0}} (\nabla^\perp\Phi-\nabla (\Delta^{-1}(\frac{f \Pi}{\Pi_{0}})))\right)
+
\left(\nabla\phi'+\nabla^\perp (\Delta^{-1}( \frac{f \pi'}{\Pi_{0}})\right)\cdot
\left(\nabla \Phi+\nabla^\perp(\Delta^{-1}(\frac{f \Pi}{\Pi_{0}}))\right)
\nonumber\\
&&\ \ \ \ \ \ \ \ \ + \frac{g}{\rho} 
    \int^{\rho}\int^{\rho_2} \frac{\pi'}{\rho} \, d\rho_1 \,
    d\rho_2  = 0 \, .\nonumber\\ \label{PrimitiveEquationHAMTWO}
\end{eqnarray}
Remarkably,  these equations are indeed Hamiltonian, 
with the Hamiltonian given by 
\begin{eqnarray}
  \H = \frac{1}{2}\int \Bigg[ (\Pi_{0}+\Pi + \pi') \, 
  \left|\nabla \phi'+  \nabla^{{\perp}}\Delta^{-1}(\frac{f \pi'}{\Pi_{0}} )  \right|^2 - g
  \left|\int^{\rho} \frac{\pi'}{\rho_1} \, d\rho_1\right|^2  \nonumber\\  
+2\pi'\left(\nabla \phi'+\nabla^\perp \Delta^{-1}(\frac{f \pi'}{\Pi_{0}})\right)\cdot
        \left(\nabla\Phi+\nabla^\perp\Delta^{-1}(\frac{f \Pi}{\Pi_{0}})\right)  
\Bigg] d\r \, .\nonumber\\
  \label{MainHamiltonianRotations}
\end{eqnarray}
Please compare this with (\ref{MainHamiltonian}). 

We now can rewrite the Hamiltonian (\ref{MainHamiltonianRotations}) in the following form:
\begin{eqnarray}
\H &=& \H_{\rm linear} +\H_{\rm nonlinear},\nonumber\\ 
\H_{\rm linear} & = &  \frac{1}{2}\int \left[ \Pi_{0} 
  \left|\nabla \phi'+  \nabla^{{\perp}}\Delta^{-1}(\frac{f \pi'}{\Pi_{0}} )  \right|^2  -g
  \left|\int^{\rho} \frac{\pi'}{\rho} \, d\rho_1\right|^2\right]d\r,
 \nonumber\\ 
 \H_{\rm  nonlinear}& =&  \H_{\rm  local}+ \H_{\rm  sweeping} + \H_{\rm  density},\nonumber\\
 \H_{\rm  local} &=&   \frac{1}{2}\int \pi'
  \left|\nabla \phi'+  \nabla^{{\perp}}(\Delta^{-1}\frac{f \pi'}{\Pi_{0}} )  \right|^2 d\r, \nonumber\\
\H_{\rm  density} &=&   \frac{1}{2}\int \Pi  
\left|\nabla \phi'+  \nabla^{{\perp}}(\Delta^{-1}\frac{f \pi'}{\Pi_{0}} )  \right|^2 d\r, \nonumber\\
 \H_{\rm  sweeping}& =& \int \pi'\left(\nabla \phi'+\nabla^\perp (\Delta^{-1}\frac{f \pi'}{\Pi_{0}} )\right)\cdot
        \left(\nabla\Phi+\nabla^\perp(\Delta^{-1}\frac{f \Pi}{\Pi_{0}}))\right)
 d\r. \nonumber \end{eqnarray}
Making the Fourier transformation and making the Boussinesq approximation
allows us to rewrite this in the 
form that generalizes (\ref{FourierHam}): 
\begin{eqnarray}
\H &=& \H_{\rm linear} +\H_{\rm nonlinear},\nonumber\\ 
\H_{\rm linear} &=& \frac{1}{2}\int d\p
\left( \Pi_{0} k^2 |\phi'_{\p}|^2 + 
\left(\frac{f^2}{k^2\Pi_{0}} -\frac{g}{\rho_0^2
  m^2} \right)|\Pi_{\p}|^2 \right)\ ,
\nonumber\\ 
\H_{\rm   nonlinear}& =& \H_{\rm  local}+ \H_{\rm  sweeping} + \H_{\rm  density}, \nonumber\\
 \H_{\rm  local} &=& \frac{1}{2}{ \frac{1}{(2\pi)^{3/2}}} 
\int d \p_1 d \p_2 d \p_3
\delta(\p_1+\p_2+\p_3)\times\left( -\K_2\cdot\K_3 \pi'_{\p_1} \phi'_{\p_2}\phi'_{\p_3}\right. 
\nonumber \\ && \left. \ \ \ \ \ \ \  \ \ \ \ \ \ \  \ \ \ \ \ \ \  \ \ \ \
- (\frac{f}{\Pi_{0}})^2  \frac{\K_2\cdot\K_3}{k_2^2
  k_3^2}\pi'_{\p_1}\pi'_{\p_2}\pi'_{\p_3} {\bf -} 
 2 \frac{f}{\Pi_{0}} \frac{\K_2\cdot\K_3^{{\perp}}}{k_3^2}\pi'_{\p_1}\phi'_{\p_2}\pi'_{\p_3}\right)\, ,
\nonumber\\
\H_{\rm  density} & = &  \frac{1}{2} { \frac{1}{(2\pi)^{3/2}}} 
\int d \p_1 d \p_2 d \p_3
\delta(\p_1+\p_2+\p_3)\left( -\K_2\cdot\K_3 \Pi_{\p_1} \phi'_{\p_2}\phi'_{\p_3}
\right. \nonumber  \\ && \left.   \ \ \ \ \ \ \ \ \ \ \ \ \ \ \ \ \ \ \ \ \ 
- (\frac{f}{\Pi_{0}})^2 \frac{\K_2\cdot\K_3}{k_2^2  k_3^2}
\Pi_{\p_1}\pi'_{\p_2}\pi'_{\p_3} {\bf +}
2 \frac{f}{\Pi_{0}} \frac{\K_2\cdot\K_3^{{\perp}}}{k_3^2}\Pi_{\p_1}\phi'_{\p_2}\pi'_{\p_3}\right)\,,
\nonumber\\ 
\H_{\rm  sweeping}& = & {\frac{1}{(2\pi)^{3/2}}} 
\int d \p_1 d \p_2 d \p_3
\delta(\p_1+\p_2+\p_3)\times \left( 
-\K_2\cdot\K_3 \pi'_{\p_1} \phi'_{\p_2}\Phi_{\p_3}\right.
\nonumber\\ && \left. \hskip -2cm
{\bf +}\frac{f}{\Pi_{0}}   \frac{\K_2\cdot\K_3^\perp}{ k_3^2}
     \pi'_{\p_1} \phi'_{\p_2}\Pi_{\p_3}
{\bf + }\frac{f}{\Pi_{0}}\frac{\K_2^\perp\cdot\K_3}{k_2^2}\pi_{\p_1}'\pi'_{\p_2}\Phi_{\p_3}
{\bf -} (\frac{f}{\Pi_{0}})^2 
\frac{\K_2\cdot\K_3}{k_2^2 k_3^2}\pi'_{\p_1}\pi'_{\p_2}\Pi_{\p_3}\right)\, ,
\nonumber\\
\label{FourierHamRotations}
\end{eqnarray}

The next step is to substitute formulas for $\Pi_\q$ and $\Phi_\q$
from (\ref{TransformationToActionVariables}), and to perform the steps
and approximations that are developed in sections (\ref{S1},\ref{S2}).
Results are the generalization of the (\ref{ResultingHamiltonian}), 
and (\ref{KernelSweep}) and (\ref{KernelDensity}):
\BEA
\H &=&\int  d \p_1 d \p_2 A^f(\p_1, \p_2) a_{\p_1}a^*_{\p_2}, 
\nonumber\\
&{\rm with }& \
\nonumber\\
A^f(\p_1,\p_2)&=& \omega_{\p_1}\delta(\p_1-\p_2) + A_{\rm  sweeping}(\p_1, \p_2)+ 
A_{\rm density }^f(\p_1, \p_2), \nonumber\\
A_{\rm  sweeping}^f(\p_1, \p_2)&  =&\frac{1}{(2\pi)^{3/2}}
\left(
 \frac{1}{2}i 
\left({\K_1-\K_2}\right)\cdot\left( \K_1+\K_2\right)
{\bf +} \frac{f}{\Pi_{0}} \K_1\cdot\K_2^\perp(\frac{1}{k_2^2} {\bf -} \frac{1}{k_1^2})
\frac{g k_1 k_2}{2\sqrt{\omega_1\omega_2}N^2}\right)
 \Phi_{\p_2-\p_1}\nonumber\\
&+&\left(
 \frac{{\bf -} i \frac{f}{\Pi_{0}} \K_1^\perp\cdot \K_2}{|\K_2-\K_1|^2}
-(\frac{f}{\Pi_{0}})^2 \frac{(\K_2-\K_1)}{({\K_1-\K_2})^2}\cdot(\frac{\K_2}{k_2^2}+\frac{\K_1}{k_1^2})\frac{g k_1 k_2}
{2\sqrt{\omega_1\omega_2}N^2}
\right)
\Pi_{\p_2-\p_1}, \nonumber\\
A_{\rm density}^f(\p_1, \p_2)& =& \frac{1}{(2\pi)^{3/2}}
\left(
\frac{N^2}{2 g} \sqrt{  \omega_{\p_1}\omega_{\p_2}} \frac{\K_2\cdot\K_3}{k_2 k_3}
+ 
\frac{i}{2}\frac{g}{N^2} \frac{f^2}{ \Pi_0^2 \sqrt{  \omega_{\p_1}\omega_{\p_2}}}
\right) \Pi_{\p_1-\p_2}.\nonumber\\
\label{NewRotationsResultingHamiltonian}
\EEA

Using these equations and repeating steps above and discarding terms proportional to $q^2$ leads to
\begin{eqnarray}
  \omega(\p,\r)&=& \OMEGAP 
-\K\cdot\nabla\Phi(\r,t)-
\frac{f}{\Pi_{0}}\K\cdot\nabla^\perp\Delta^{-1}\Pi(\r,t)
\nonumber\\&&
+\frac{N^2}{2 g}\Pi(\r,t)\sigma_{\bf p}
\nonumber\\&+&\frac{f^2 g }{\Pi_0^2\omega_p N^2}\Pi(\r,t)
+\frac{f^2 g}{2\omega_\p \Pi_0^2 N^2}\int \cos({2\theta_{\q\p}})\Pi(\q,t) e^{i \r\cdot \q}d q.
  \nonumber\\
\label{FinalEquationTookMeTwoMonthsToDerive}
\end{eqnarray}
Here $\theta_{\q\p}$ is an angle between the horizontal part of wave vector ${\bf q} = \p_1 - \p_2 $ (\ref{Wigner}) and the horizontal part of ${\bf p}$, that is ${\bf k}$, with the sign defined using the right-hand rule going from $q$ TO $p$.  The Eulerian frequency $\OMEGAP$ is given by (\ref{InternalWavesDispersion}).   Note that the first line is simply $\OMEGAP-\K\cdot{\cal U}$, the second line is the density term that we have considered above, and the third line is the contribution from a rotating ocean.  Equation (\ref{SpaceDependentFrequency}), which is lines one and two of (\ref{FinalEquationTookMeTwoMonthsToDerive}) with $\cal{U}$$ = \nabla\Phi$, applies to a non-rotating ocean.  In a rotating ocean, a wave whose frequency is high enough not to be impacted by rotation is described using lines one and two.

\end{document}